\pdfoutput=1
\documentclass[conference,letterpaper]{IEEEtran}
\usepackage[T1]{fontenc}
\usepackage{cite}
\usepackage{booktabs}
\usepackage{pgfplots}
\pgfplotsset{compat=1.18}
\usepackage{url}
\usepackage[hidelinks]{hyperref}
\hypersetup{pdftitle={Preference Is Not Position: Evaluating Reranking on NevIR in Fixed Hybrid Candidate Pools},
pdfauthor={Yuning Bian, Lele Xu, Yifan Yuan, Xinyu Yin, Chengyan Zhan}}
\usepackage{balance}
\usepackage{eso-pic}
\newcommand{\ieeenotice}{\AddToShipoutPictureBG*{\AtPageLowerLeft{\parbox[b][0.62in][c]{\paperwidth}{\centering\scriptsize
\begin{minipage}{0.86\paperwidth}\centering
\copyright~2026 IEEE. Personal use of this material is permitted. Permission from IEEE must be obtained for all other uses, in any current or future media, including reprinting/republishing this material for advertising or promotional purposes, creating new collective works, for resale or redistribution to servers or lists, or reuse of any copyrighted component of this work in other works.
\end{minipage}}}}}

\title{Preference Is Not Position: Evaluating Reranking on NevIR in Fixed Hybrid Candidate Pools}
\author{%
\IEEEauthorblockN{Yuning Bian\textsuperscript{*}, Lele Xu\textsuperscript{*}, Yifan Yuan\textsuperscript{*}, Xinyu Yin, Chengyan Zhan}
\IEEEauthorblockA{School of Computer Science and Technology\\
Anhui University of Technology\\
Ma'anshan, China\\
bianyuning@gmail.com, jixu0090@gmail.com, yyifan355@gmail.com\\
Stressful0614@outlook.com, zhan82789@gmail.com\\
All authors are undergraduate students.\\
\textsuperscript{*}~These authors are co-first authors.}}

\begin{document}
\maketitle
\ieeenotice

\begin{abstract}
Near-identical passages can differ in a decisive condition, such as negation. A ranker may order such a pair correctly while placing the preferred passage far down the retrieved list. We examine this distinction using official NevIR pairs jointly covered by fixed candidate pools from dense, learned sparse, and BM25 retrieval. On 2,736 Test query directions, a pairwise-trained ranker improves strict per-query preference over fixed fusion (56.62\% versus 52.30\%) while reducing preferred-passage mean reciprocal rank (MRR) from 0.5764 to 0.3055. An exploratory fixed-model replay of a variant with background sampling and revised supervision yields MRR of 0.5859 and strict preference of 54.53\%. Bounded pointwise reranking of the fusion top~20 improves both outcomes. Qwen3-14B-AWQ with fusion tie-breaking reaches 77.45\% strict preference, 58.99\% NevIR-style paired accuracy across 1,363 complete pairs, and preferred-passage MRR of 0.7509. A BGE cross-encoder reaches 69.77\% strict preference and MRR of 0.7404. A post-hoc analysis finds that the branch where equal Qwen grades for the two designated passages trigger fusion tie-breaking accounts for 27\% of evaluated directions and contains most remaining strict-preference failures. These findings support joint evaluation of pairwise preference, designated-passage position, and computational cost, while distinguishing model judgments from the combined ranking procedure.
\end{abstract}


\section{Introduction}
\label{sec:introduction}

Retrieval can select passages about the right topic while missing a condition that determines which passage answers a query. Negation makes this failure particularly visible: two nearly identical passages can support opposite answers~\cite{weller-etal-2024-nevir}. After multiple retrievers contribute candidates, a reranker must both prefer the passage that satisfies the query and place it early enough to be used. These requirements motivate a joint examination of local preference and position in the returned list.

NevIR introduced contrastive passage pairs to diagnose negation sensitivity; its paired accuracy requires correct preferences for both associated queries~\cite{weller-etal-2024-nevir}. NevIR and its reproduction already show that improvements on negation-sensitive evaluation need not translate into improvements on broader retrieval benchmarks~\cite{weller-etal-2024-nevir,elsen_reproducing_2025}. We ask a more specific question within one evaluation environment: \emph{when the candidate set is fixed, do changes that improve the designated pairwise preference also improve the preferred passage's position among surrounding candidates?} A correct preference alone cannot answer this question: ranking the preferred passage 100th and its counterpart 101st satisfies the local ordering while leaving both far down the list.

We study this distinction in fixed pools formed by the union of dense, learned sparse, and BM25 retrieval results. All methods use the same retrieved passages, without additional retrieval or access to source text beyond those passages. We measure strict per-query preference for the officially designated preferred passage and its reciprocal rank in the full candidate list. Both metrics are conditional on pair coverage, which is reported separately.

On Test, the pair-trained ranker improves strict preference over fusion but lowers preferred-passage MRR. Background sampling with revised supervision recovers MRR while reducing strict preference relative to the pair-trained ranker; these joint changes do not isolate the cause of recovery.

We next examine whether bounded pointwise reranking improves both outcomes relative to fixed fusion. BGE and Qwen score the same fusion top-20 window, with fusion tie-breaking and an unchanged tail. We then use a post-hoc analysis of Qwen outputs to examine whether the combined ranking performance reflects distinctions expressed by the scoring model.

On Test, Qwen with fusion tie-breaking raises strict preference from 52.30\% to 77.45\% (25.15 points, source-group 95\% interval [23.54, 26.78]) and preferred-passage MRR from 0.5764 to 0.7509; a BGE cross-encoder in the same window reaches 69.77\% and 0.7404. Section~\ref{sec:cost} reports their separately recorded costs and timing scopes. A post-hoc analysis finds that 27\% of directions rely on fusion to resolve equal Qwen grades, mostly at the highest grade; this branch contains 61\% of remaining strict-preference failures. The study contributes a within-pool diagnosis of the separation between pairwise preference and designated-passage position, an evaluation of training and bounded reranking interventions against both outcomes with cost, and a characterization of tied grades and their role in the remaining errors.

\section{Related Work}
\label{sec:related}

NevIR pairs near-identical passages whose relevance is reversed by negation and scores a system only when both associated queries are ordered correctly~\cite{weller-etal-2024-nevir}. Its reproduction extends the comparison to newer rerankers and LLM-based rankers and confirms that pairwise success on this benchmark does not follow general retrieval quality~\cite{elsen_reproducing_2025}. Their NevIR evaluations use the designated pairs, while broader retrieval quality is assessed separately. We instead measure pairwise preference and designated-passage position within the same retrieved NevIR candidate pools.

Rank fusion~\cite{cormack2009rrf} and normalized-score fusion~\cite{bruch2023fusion} combine signals from multiple retrievers. Here, the retrieved union fixes candidate membership, and a frozen weighted score combination determines the reference order.

Gradient-boosted LambdaMART~\cite{burges2010lambdamart,ke2017lightgbm} remains a common feature-based reranker. Pointwise neural rerankers score each query--passage pair independently~\cite{nogueira-etal-2020-document}; instruction-following LLMs have been used as listwise~\cite{sun-etal-2023-chatgpt}, pairwise~\cite{qin-etal-2024-large}, and pointwise~\cite{zhuang-etal-2024-beyond} rerankers over a first-stage top-$k$. Reranking a bounded window is therefore not new. Our contribution is the evaluation design: on one fixed pool, we ask whether training-based and window-based interventions move designated-pair preference and designated-passage position in the same direction.

Zhuang et al.~\cite{zhuang-etal-2024-beyond} compare generated grades, including a 0--4 scale, with likelihood-derived scores and discuss ties as a limitation of generated grades. Our analysis characterizes the frequency of ties between designated passages, their concentration at the highest grade, and their share of remaining errors within the fixed NevIR candidate pools.

\section{Setting and Evaluation}
\label{sec:setting}

Our first question (RQ1) is whether improvements in designated-pair preference coincide with improvements in designated-passage position under training interventions. RQ2 asks whether bounded text reranking improves both strict preference and preferred-passage position relative to fixed fusion, and documents the associated execution costs. For each query $q$, we fix the deduplicated union $C_q$ of dense, learned sparse, and BM25 results. All comparisons reuse the saved query, passage text, and route scores and ranks. Missing designated passages are not inserted. The saved route depths are 150/50/100, respectively. Dense retrieval uses \texttt{text-embedding-v4} from the Qwen3-Embedding series (1,024 dimensions), learned sparse retrieval uses \texttt{doubao-embedding-vision-251215}, and BM25 uses the project's SQLite FTS5 implementation.

We use the official passage preferences of NevIR~\cite{weller-etal-2024-nevir}. The training/development/confirmation retrieval corpus contains 1,761 distinct passages from its Train and Validation material, with one passage per document. Development contains 38 pairs; confirmation contains the remaining 187 Validation pairs. In these partitions, source groups connect examples through source identifiers, identical passages, or queries matching exactly or by normalized token sequence or multiset. Table~\ref{tab:populations} separates planned queries from usable directions. Of 1,871 jointly covered training directions, two with identical queries and opposing preferences are excluded, leaving 1,869 for training. Development was used for parameter selection and early stopping. Confirmation was first used for model comparison, then reused for diagnostics and reranking. An exploratory GPT-6 pointwise-reranking pilot on development and confirmation motivated the subsequent evaluation of open-weight rerankers; confirmation was not a fresh holdout for every intervention.

\begin{table}[t]
\centering
\caption{Data populations. Used queries and source groups refer to loss-eligible training directions or jointly covered evaluation directions, after excluding structural conflicts where applicable. Complete pairs require both directions.}
\label{tab:populations}
\small
\begin{tabular}{lrrrr}
\toprule
Partition & Planned & Used & Complete & Source\\
 & queries & queries & pairs & groups\\
\midrule
Train & 1,896 & 1,869 & -- & 472\\
Development & 76 & 74 & 37 & 19\\
Confirmation & 374 & 371 & 185 & 96\\
Test & 2,766 & 2,736 & 1,363 & 699\\
\bottomrule
\end{tabular}
\end{table}

Let $d_q^+$ denote the officially preferred passage and $d_q^-$ its counterpart. Let $Q_c$ contain queries with eligible official labels and both passages in $C_q$, excluding structural conflicts. Primary strict preference is the fraction of $Q_c$ with $d_q^+$ scored strictly above $d_q^-$. Composite rerankers use the ordered scoring keys in Section~\ref{sec:methods}. Equal keys remain ties; passage identifiers determine list positions but never strict successes. Bidirectional accuracy requires strict correctness in both directions of complete pairs, following NevIR's paired-accuracy aggregation~\cite{weller-etal-2024-nevir}. Our fixed-pool protocol allows us to evaluate designated-pair preference and preferred-passage position within the same retrieved lists. Comparisons use this shared setting; coverage conditioning and composite ranking preclude direct comparison with published given-pair NevIR scores. Coverage concerns the full pool, not the reranking window. Preferred-passage MRR is $|Q_c|^{-1}\sum_{q\in Q_c}1/r_q^+$, with $r_q^+$ the one-based full-pool rank. Preferred@$k$ is the percentage of query directions in $Q_c$ for which $r_q^+ \leq k$. We also report both designated passages' mean ranks and the count with both in the top 10. None of these position measures assesses the relevance of unjudged background candidates.

Official preferences define the main analysis. A later human review of development retained 19 ties and four unresolved judgments among 76 queries. Of 53 strict human preferences, 52 have joint candidate coverage: 47 agree with the official direction and five disagree. This is a label-agreement check on already inspected data, not a replacement gold standard. In the window-reranking evaluation, unavailable scores trigger fusion fallback without removing queries from the evaluation. Cost accounting covers all planned queries and distinguishes judgments, requests, tokens, and wall-clock time. Where uncertainty intervals are reported, resampling uses source groups rather than treating associated queries as independent; the list-diagnostic aggregates below are descriptive.

Test uses a separate corpus of 2,081 deduplicated passages from the official Test split, one passage per document. Four also occur in Train/Validation and are retained. Of 2,766 planned directions, 2,738 jointly cover the designated passages; excluding two structural conflicts leaves 2,736 directions and 1,363 complete pairs. Ten eligible directions from incomplete pairs enter per-query metrics but not paired accuracy. Reported bootstrap intervals use 699 source-identifier-prefix groups. Earlier E0/fusion/background-training Test aggregates were received and arithmetically checked before Qwen/BGE evaluation; that analysis removed identifier-based tie breaking from strict correctness. Qwen/BGE configurations were fixed before their Test scoring. All 78 primary-population fallback queries remain in the denominator.

\section{Comparators and Interventions}
\label{sec:methods}

\textit{Fixed fusion.} The reference combines normalized dense, sparse, and BM25 scores with weights 0.70, 0.15, and 0.15. Fusion preprocessing filters route scores at 0.30/0.20/0, applies $\log(1+s)$ to sparse/BM25 scores, and min--max normalizes within each retained route; a constant-score route receives ones. Weights are renormalized over nonempty routes. Missing contributions, including passages with no surviving route score, are zero. Filtering affects fusion contributions, not membership of the saved union. Weights, thresholds, and depths are inherited unchanged from the project's production hybrid configuration, which was set on an unrelated Chinese corpus; nothing was adjusted for NevIR. These saved scores order the complete pool and supply both text rerankers' top-20 window.

\textit{Pair-trained E0.} The English comparator is a LightGBM LambdaMART ranker~\cite{ke2017lightgbm} with 38 features computed from the complete candidate pool, including route scores and ranks, route agreement, and lexical matching signals. Only the two designated passages enter each training query's loss, labeled 1/0; background candidates affect pool features and inference but receive no training labels. E0 serves as a diagnostic comparator for pair-restricted supervision applied to full-pool ranking. E0 inherits the previously selected configuration with seven leaves, depth three, learning rate 0.03, and L2 regularization 1. E0 is trained once with a fixed random seed, up to 300 boosting rounds, and early-stopping patience of 40 rounds. Source-balanced weights and stopping on development strict preference averaged within and then across source groups yield 67 trees. Both training arms retain LambdaRank truncation level 2. Parameters were not selected by the confirmation comparison reported here.

\textit{Background training ($N=8$).} This variant retains E0's features, hyperparameters, seed, and development stopping criterion. Each training query adds eight candidates sampled without replacement from its pool outside the designated pair, with a deterministic per-query seed recorded in the repository. Labels become 2/1/0 for the preferred passage, counterpart, and sampled background, with gains $[0,1,2]$. The loss grows from 3,738 to 18,690 rows over the same 1,869 queries; the fitted model has 69 trees. Background labels are weak assumptions, not human judgments of irrelevance. Sampling, label levels, and gains change together, making this a joint training intervention rather than an isolated test of background count.

\textit{Shared text-reranking procedure.} Both BGE and Qwen score query--passage inputs separately within the same fusion top 20. The window size was fixed during the initial GPT pilot, before any Qwen or BGE scoring, because the fusion top 20 covered both designated passages for 72/74 development and 355/371 confirmation directions; it was not tuned on Test. Neither scorer receives official pair identities, labels, or other candidates. The head is sorted by text score, then fusion score, then passage identifier; the tail keeps its fusion order. Text grades take priority within the window, and fusion scores break grade ties. Score interpolation was not evaluated. On successful scoring, strict comparisons use the key $(1,s_{\mathrm{text}},s_{\mathrm{fusion}})$ in the head and $(0,0,s_{\mathrm{fusion}})$ in the tail. Identifier ordering is excluded from strict comparisons. Any unavailable head score restores both the original fusion order and raw fusion scores for strict comparisons, retaining that query in the evaluation. An auxiliary Qwen variant replaces fusion with E0 for grade tie-breaking, leaving Qwen scores, window membership, tail order, and fallback unchanged. Pair-only scoring of designated passages is an auxiliary diagnostic and does not replace this candidate-window evaluation; on Test, Qwen's pair-only and window stages use separate requests, whereas BGE scores the deduplicated input union once.

\textit{Scoring models.} BGE uses \texttt{bge-reranker-v2-m3}~\cite{bgeRerankerCard} through the Transformers sequence-classification interface in FP16, returning raw logits without discretization. No input was truncated. New Qwen requests and the separate repeat use \texttt{Qwen3-14B-AWQ} in thinking mode~\cite{yang2025qwen3}, temperature 0, an 8,192-token context, and a 6,144-token output cap. The fixed prompt requests an integer score from 0 to 4: unrelated, condition contradicted or misattributed, required evidence absent, satisfied with an implicit detail, or explicitly satisfied. We parse the generated grade directly; this scoring approach and likelihood-based alternatives are both evaluated by Zhuang et al.~\cite{zhuang-etal-2024-beyond}. The generated grades make the scorer's stated condition satisfaction explicit and allow grade ties to be analyzed separately from the fusion ordering.

The 14B configuration was selected on development before the first Qwen confirmation run; a later 8B development comparison scored slightly higher on top-20 reranking. The first Qwen confirmation run reused valid scores potentially generated under earlier output limits and allowed internal retries. The separate repeat used an empty cache and no failure retries, with seven of 374 queries falling back; it changed the confirmation strict count from 280 to 277 of 371, the only run-to-run variability measured. First-run scores supply the confirmation cost analysis. Each Test configuration was run once.

\begin{samepage}
\textit{Materials.} A standalone CPU package provides evaluation code, saved scores, frozen configurations, and the grading prompt, regenerating the three tables, two figure data sets, and auxiliary tie-breaking results from cached scores.\footnote{\url{https://github.com/ql-link/LinkRag-Eval/releases/tag/reproduction}}
\par
\end{samepage}

\section{Results}
\label{sec:results}

Table~\ref{tab:main} compares preference and position for five rankers on the primary Test population; Section~\ref{sec:cost} reports costs.

\begin{table*}[t]
\centering
\caption{Test results on the fixed candidate pools ($n=2{,}736$ directions; 1,363 complete pairs). Strict: per-query strict preference. Paired: NevIR-style bidirectional accuracy. Pref@$k$ and MRR: position of the officially preferred passage in the full pool. Ties: query directions with equal final scoring keys for the two designated passages, before identifier ordering. Fallback: queries restored to fusion order after unavailable window scoring. $\Delta$: strict difference versus fixed fusion in percentage points with a source-group bootstrap 95\% interval. The Qwen contrast is primary; the BGE contrast is auxiliary.}
\label{tab:main}
\small
\setlength{\tabcolsep}{3pt}
\begin{tabular}{lrrrrrrrrr}
\toprule
 & \multicolumn{2}{c}{Preference (\%)} & \multicolumn{4}{c}{Position of $d_q^+$ ($d_q^-$)} & & & \\
\cmidrule(lr){2-3}\cmidrule(lr){4-7}
Ranker & Strict & Paired & @1 (\%) & @3 (\%) & MRR & Mean rank & Ties & Fallback & $\Delta$ strict [95\% CI] \\
\midrule
Fixed fusion & 52.30 & 10.79 & 33.59 & 86.18 & 0.5764 & 3.96 (4.16) & 7 & -- & -- \\
E0 (pair-trained LTR) & 56.62 & 21.20 & 28.18 & 31.36 & 0.3055 & 92.29 (106.61) & 29 & -- & -- \\
$N=8$ background LTR & 54.53 & 15.04 & 34.43 & 86.07 & 0.5859 & 3.61 (3.86) & 70 & -- & -- \\
Fusion top-20 + BGE & 69.77 & 43.36 & 57.02 & \textbf{95.07} & 0.7404 & \textbf{3.03} (3.79) & 5 & 0 & +17.47 [15.81, 19.05] \\
Fusion top-20 + Qwen & \textbf{77.45} & \textbf{58.99} & \textbf{59.21} & 92.76 & \textbf{0.7509} & 3.08 (4.75) & 5 & 78 & +25.15 [23.54, 26.78] \\
\bottomrule
\end{tabular}
\end{table*}

\subsection{Preference versus Position under Training Interventions (RQ1)}
\label{sec:diagnosis}

\begin{figure}[t]
\centering
\begin{tikzpicture}
\begin{axis}[
width=0.98\columnwidth,height=5.0cm,
xmin=48,xmax=82,ymin=0.2,ymax=0.85,
xtick={50,55,60,65,70,75,80},ytick={0.2,0.4,0.6,0.8},
xlabel={Strict per-query preference (\%)},
ylabel={Preferred-passage MRR},
label style={font=\small},tick label style={font=\footnotesize},
grid=major,grid style={gray!18},axis line style={gray!70},
enlargelimits=false,clip=false]
\addplot[only marks,mark=square*,mark size=2.8pt,color=blue!65!black]
coordinates {(52.3026,0.5764)};
\node[anchor=north,font=\footnotesize] at (axis cs:52.3026,0.5604) {Fusion};
\addplot[only marks,mark=triangle*,mark size=3.2pt,color=orange!80!black]
coordinates {(56.6155,0.3055)};
\node[anchor=north,font=\footnotesize] at (axis cs:56.6155,0.2895) {E0};
\addplot[only marks,mark=*,mark size=2.8pt,color=teal!80!black]
coordinates {(54.5322,0.5859)};
\node[anchor=south,font=\footnotesize] at (axis cs:54.5322,0.6019) {$N=8$};
\addplot[only marks,mark=diamond*,mark size=3.2pt,color=purple!80!black]
coordinates {(69.7734,0.7404)};
\node[anchor=north,font=\footnotesize] at (axis cs:69.7734,0.7244) {Top-20+BGE};
\addplot[only marks,mark=pentagon*,mark size=3.2pt,color=red!75!black]
coordinates {(77.4488,0.7509)};
\node[anchor=south,font=\footnotesize] at (axis cs:77.4488,0.7669) {Top-20+Qwen};
\end{axis}
\end{tikzpicture}
\caption{Strict preference and preferred-passage MRR on the same Test population ($n=2{,}736$ directions). Higher is better on both axes. Points are observed aggregates without uncertainty intervals.}
\label{fig:preference-position}
\end{figure}
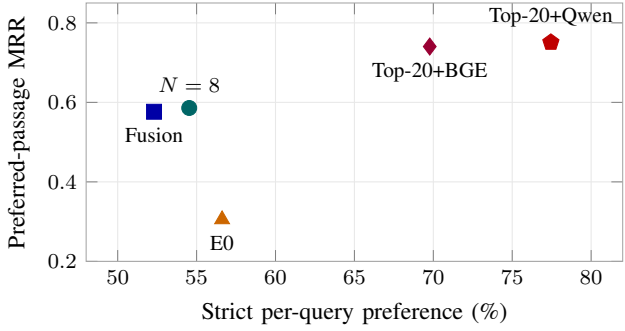

Table~\ref{tab:main} and Figure~\ref{fig:preference-position} compare the three training/fusion conditions on the 2,736 covered Test directions. E0 obtains 1,549 strict preferences (56.62\%) against fusion's 1,431 (52.30\%), yet its preferred-passage MRR is 0.3055 versus 0.5764, and the mean ranks of the preferred passage and its counterpart are 92.3/106.6 against 4.0/4.2. Both designated passages appear in the top 10 for 64 Test directions under E0 and 2,547 under fusion. E0's higher preference count therefore coexists with designated passages pushed deep into the pool; it is not evidence of better ordering of the whole pool. Evaluating only preference on these saved rankings would favor E0 over fusion while missing the position collapse. The confirmation set shows the same pattern (E0 208/371 versus fusion 189/371 strict; MRR 0.3169 versus 0.6017).

Background training moves in the opposite direction. With $N=8$, Test preferred-passage MRR rises to 0.5859, mean ranks are 3.6/3.9, and both passages occupy the top 10 for 2,609 directions, but strict correctness falls to 1,492 (54.53\%), 57 below E0, with 70 ties. Confirmation shows the same pattern of higher MRR and lower strict preference (MRR 0.3169 to 0.6133; strict 208 to 198 of 371), as does development, which was used for early stopping. Relative to fusion, $N=8$ gains 61 strict successes on Test and 9 on confirmation.

Recalculation on the saved candidate snapshot reproduced the earlier E0, fusion, and $N=8$ Test aggregates. For $N=8$, fixed-model replay generated new per-query scores; original scores and execution logs remain unavailable, leaving its pre-Test freeze timeline unverified. We therefore treat the $N=8$ replay as exploratory evidence. The results are consistent with the revised supervision encouraging designated passages to outrank sampled background. They do not establish that the sampled candidates are irrelevant or identify which component of the joint intervention drives recovery; only one background count and seed were evaluated.

These comparisons answer RQ1 at the level of the observed setting: the methods' ordering by strict preference differs from their ordering by designated-passage position. Both outcomes should therefore accompany any RQ2 comparison, together with separately measured cost.

\subsection{Bounded Text Reranking (RQ2)}
\label{sec:rq2}

The primary contrast is Qwen top-20 reranking with fusion tie-breaking against fixed fusion. Strict preference rises from 1,431 to 2,119 of 2,736 directions (52.30\% to 77.45\%), a difference of 25.15 percentage points with a source-group bootstrap 95\% interval of [23.54, 26.78] (2,000 resamples over 699 groups). Bidirectional accuracy rises from 147 to 804 of 1,363 pairs (10.79\% to 58.99\%). At the query-direction level, the strict-preference change comprises 753 corrections and 65 regressions relative to fusion; net gain is positive in 502 source groups, zero in 175, and negative in 22. Preferred-passage MRR rises from 0.5764 to 0.7509 and preferred@1 from 33.59\% to 59.21\%, so this intervention improves both strict preference and preferred-passage MRR relative to fusion. Against E0, the strict difference is 20.83 points [18.72, 22.91]. Regrouping by the Train/Validation rules (697 groups) leaves the Qwen--fusion interval essentially unchanged ([23.55, 26.76]).

The same window with BGE scores reaches 69.77\% strict preference and 43.36\% paired accuracy. The strict-preference gain over fusion is 17.47 points [15.81, 19.05], with 788 corrections and 310 regressions. BGE corrects more fusion errors than Qwen (788 versus 753), but also incurs more regressions (310 versus 65). At the aggregate level, the Qwen procedure's higher final strict preference therefore reflects fewer regressions rather than more corrections. BGE has higher preferred@3 and a lower preferred-passage mean rank, with no fallback. The 7.68-point strict difference between them is descriptive; no Qwen--BGE test was pre-specified.

\begin{table}[t]
\centering
\caption{Test designated-pair-only scoring (L1) versus top-20 reranking (L2; Table~\ref{tab:main}). L1 counts ties and unavailable directions as failures. Unav.: query directions with at least one unavailable designated-passage score. Qwen uses separate L1/L2 scores; BGE reuses shared-input logits. E0 is repeated as a full-pool reference.}
\label{tab:l1l2}
\small
\setlength{\tabcolsep}{2pt}
\begin{tabular}{lrrrrrr}
\toprule
 & \multicolumn{4}{c}{L1: designated pair} & \multicolumn{2}{c}{L2: top-20} \\
\cmidrule(lr){2-5}\cmidrule(lr){6-7}
Scorer & Strict (\%) & Paired (\%) & Ties & Unav. & Strict (\%) & Paired (\%) \\
\midrule
E0 & 56.62 & 21.20 & 29 & 0 & 56.62 & 21.20 \\
Qwen & 64.80 & 48.20 & 777 & 23 & 77.45 & 58.99 \\
BGE & 69.96 & 43.43 & 2 & 0 & 69.77 & 43.36 \\
\bottomrule
\end{tabular}
\end{table}

Table~\ref{tab:l1l2} contrasts pair-only scoring with window reranking. Pair-only strict preference favors BGE over Qwen (69.96\% versus 64.80\%), whereas window reranking favors the Qwen procedure (77.45\% versus 69.77\%). Their ordering by strict preference therefore does not carry over from pair-only scoring to window reranking. Qwen's pair-only outputs contain 777 grade-tied directions (28.4\%) and 23 directions with an unavailable designated-passage score, all counted as strict-preference failures; BGE's logits tie in two directions. This descriptive difference does not isolate fusion tie-breaking from separately obtained scores, window membership, or fallback; Section~\ref{sec:discussion} therefore analyzes the actual L2 outputs.

Window coverage and fallback limit where reranking can affect preference. In the primary population, both designated passages lie within the fusion top 20 for 2,629 of 2,736 directions. The remaining 107 directions and the 78 fallback queries overlap in one direction. Their union comprises 184 directions whose designated preference is unchanged from fusion, contributing 93 strict successes.

\subsection{Cost}
\label{sec:cost}

Both rerankers score at most 20 passages per query in the window stage. On one RTX 4090, BGE scored the 55,502 unique query--passage inputs in the union of the pair-only and window stages in 366 s of inference. E0's local CPU run, including model loading, feature computation, and output, took 73 s for all 2,766 queries; the original $N=8$ scoring time was not recorded.

On one RTX 4090, Qwen in thinking mode produced 60,724 recorded responses across the pair-only and window stages, with recorded usage of 38.7 M tokens (21.5 M input, 17.2 M output). The recorded execution intervals for the window stage sum to about 6.5 h (23.3 ks); service setup and data transfer are excluded. Across all planned queries, 79 window inputs were unavailable, and usage records were missing for eight inputs. BGE inference time and Qwen's elapsed intervals cover different workloads and timing scopes; batching and concurrency also differ, so these are execution records rather than a matched-workload latency comparison.

On the already-exposed confirmation set, cached-score replay yields 274/371 strict successes at $K=3$ versus 280/371 at $K=20$, using 2.99 effective judgments per planned query. Two query-level rules---one based on the top-two fusion-score margin and the other on route-winner disagreement or a missing route---yield 275 and 267 successes, with 4.5\% and 13.4\% fewer judgments than full-call $K=20$, respectively. These replays compare strict preference and judgment counts only; they do not establish preservation of preferred-passage position or a latency improvement.

\section{Discussion and Future Work}
\label{sec:discussion}

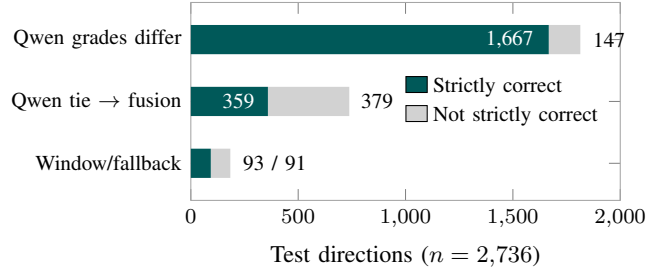
\begin{figure}[t]
\centering
\begin{tikzpicture}
\begin{axis}[
xbar stacked,
width=0.82\columnwidth,height=4.2cm,
xmin=0,xmax=2000,ymin=0.4,ymax=3.6,
xlabel={Test directions ($n=2{,}736$)},
ytick={1,2,3},
yticklabels={Qwen grades differ,Qwen tie $\rightarrow$ fusion,Window/fallback},
y dir=reverse,
label style={font=\small},tick label style={font=\footnotesize},
bar width=11pt,
legend style={font=\footnotesize,at={(0.98,0.5)},anchor=east,draw=none,fill=none},
legend cell align=left,legend columns=1,
axis line style={gray!70},
clip=false,
after end axis/.code={
\node[font=\footnotesize,anchor=east,text=white] at (axis cs:1640,1) {1,667};
\node[font=\footnotesize,anchor=west] at (axis cs:1830,1) {147};
\node[font=\footnotesize,anchor=east,text=white] at (axis cs:340,2) {359};
\node[font=\footnotesize,anchor=west] at (axis cs:750,2) {379};
\node[font=\footnotesize,anchor=west] at (axis cs:200,3) {93 / 91};
}]
\addplot[fill=teal!70!black,draw=none] coordinates {(1667,1) (359,2) (93,3)};
\addplot[fill=gray!35,draw=none] coordinates {(147,1) (379,2) (91,3)};
\legend{Strictly correct,Not strictly correct}
\end{axis}
\end{tikzpicture}
\caption{Which signal decided each Test direction under Qwen top-20 reranking. Branches follow the actual resorting rule and are mutually exclusive. Conditional accuracy is 91.90\%, 48.64\%, and 50.54\% from top to bottom. All 753 corrections and 65 regressions relative to fusion lie in the top branch; 571 of the 738 tied directions have both passages graded~4. Qwen-grade ties are distinct from final-key ties in Table~\ref{tab:main}.}
\label{fig:decomposition}
\end{figure}

In a post-hoc analysis of the saved Qwen L2 outputs (Figure~\ref{fig:decomposition}), the within-window grade-tie branch covers 738 of 2,736 directions (26.97\%) but contains 379 of 617 final non-successes (61.43\%). Retaining the fusion relation within this branch follows from the tie-breaking rule; the empirical finding is its frequency and concentration of residual errors. The different-grade branch contains 147 non-successes among 1,814 directions, yielding 91.90\% conditional accuracy. These groups are defined by observed outputs, not calibrated confidence.

Most within-window ties are at the highest grade: both passages receive 4 in 571/738 tied directions (77.37\%). In the tie branch, 285 of the 379 non-successes occur among double-4 outputs. The prompt defines 4 as explicitly satisfying all conditions. Assuming the official preferences are valid, these double-4 outputs fail to express the required distinction. These outputs exhibit highest-grade saturation under a rubric that asserts satisfaction, rather than uncertainty. The aggregate does not identify which condition was missed or establish score granularity as its cause.

Changing only the grade tie-breaker from fusion to E0 raises strict preference from 77.45\% to 78.47\%, but lowers preferred-passage MRR from 0.7509 to 0.7148. With Qwen outputs and window membership held fixed, this auxiliary comparison shows that preference--position divergence also occurs within bounded reranking.

The concentration of residual non-successes in the grade-tie branch motivates controlled comparisons of alternative scoring rules or additional pairwise disambiguation for tied candidates. Prior experiments show that adding more rating levels does not consistently improve ranking~\cite{zhuang-etal-2024-beyond}. The effects of these alternatives on corrections, new errors, preferred-passage position, and inference cost remain untested in our setting. They require prospective evaluation rather than selection on this Test set; no aggregate here establishes improved semantic understanding.

\balance
\section{Limitations}
\label{sec:limitations}

\textit{Labels.} A single-annotator, model-assisted review of 122 error-selected cases raised 55 label disputes, all from confirmation. These judgments did not alter the official labels or scores. The selected cases cannot estimate the overall error distribution, and the disputes do not establish benchmark-label errors.

\textit{Contamination and wording.} NevIR has been public since 2023, so exposure of Qwen3 or BGE to it cannot be excluded. A small paraphrase probe on twelve scenarios (24 directions per version) yielded 23, 20, and 23 strict Qwen successes for the original, synonym, and entity-renamed versions; the losses in the synonym version were ties. This is a wording-sensitivity observation, not a membership test.

\textit{Scope.} Results concern English NevIR pairs under one fixed retrieval configuration. Comparisons are between complete procedures and do not isolate model choice or reordering scope. Extending these findings to natural query distributions, other languages, retrieval settings, or downstream answer quality requires separate evaluation.

\section{Conclusion}
\label{sec:conclusion}

In this fixed-pool evaluation, E0 achieved a higher observed strict-preference rate than fusion but a lower preferred-passage MRR. The $N=8$ variant achieved a higher MRR and a lower strict-preference rate than E0. Thus, the two metrics ranked the evaluated procedures differently. Bounded pointwise reranking of the fusion top 20 improved both outcomes on Test: 77.45\% strict preference and 0.7509 preferred-passage MRR with Qwen3-14B-AWQ, against 52.30\% and 0.5764 for fusion, using a reranking window of at most 20 passages per query. BGE reached 69.77\% strict preference and MRR of 0.7404; the different timing scopes preclude a like-for-like speed comparison. The result belongs to the combined procedure: replacing fusion with E0 for grade tie-breaking increased strict preference but reduced preferred-passage MRR. Equal within-window grades leave 27\% of directions to fusion, mostly with both passages graded 4; this branch contains 61\% of remaining strict-preference failures. Preference and position should be reported together, and tied judgments are the natural target for future controlled comparisons.

\section*{Acknowledgment}
The authors used ChatGPT and Claude for
code development, manuscript revision, and methodological
feedback. The authors made the final decisions and take
responsibility for this work.

\bibliographystyle{IEEEtran}
\bibliography{references}
\end{document}